\documentstyle[aps,prl]{revtex}
\input epsf.tex
\def\ltap{\ \raise.3ex\hbox{$<$\kern-.75em\lower1ex\hbox{$\sim$}}\ }
\def\gtap{\ \raise.3ex\hbox{$>$\kern-.75em\lower1ex\hbox{$\sim$}}\ }

\def\etal{{\it et. al.}}
\def\eg{{\it e.g.}}

\def\CH{{\cal H}}
\def\CO{{\cal O}}
\def\CC{{\cal C}}
\def\TeV{\,{\rm TeV}}\def\GeV{\,{\rm GeV}}
\def\be{\begin{equation}}
\def\ee{\end{equation}}
\def\nn{\nonumber}
\def\pyidk{PHY-9057135}
\begin{document}
\twocolumn[
\widetext
\vskip 10pt
{\large\bf\centering B-Factory
Physics from  Effective Supersymmetry\\}\vskip5pt

{\rm\centering Andrew G. Cohen$^{a}$, David B. Kaplan$^{b}$,
 Fran\c{c}ois Lepeintre$^{b,c}$ and Ann E. Nelson$^c$\\}
\vskip1.5pt

{\small\it\centering\ignorespaces
   (a) Department of Physics, Boston University, Boston, MA 02215, USA\\
     (b) Institute for Nuclear Theory 1550, University of Washington,
         Seattle, WA 98195-1550, USA \\
    (c) Department of Physics 1560, University of Washington,
       Seattle, WA 98195-1560, USA\\
}

\begin{abstract} We discuss how to extract non-Standard Model effects
from B-factory phenomenology.  We then analyze the prospects for uncovering
evidence for Effective Supersymmetry, a class of supersymmetric models
which naturally suppress flavor changing neutral currents and electric
dipole moments without squark universality or small CP violating
phases, in experiments at BaBar, BELLE, HERA-B, CDF/D0 and LHC-B.
\end{abstract}
\vskip.5pc

]
\narrowtext

The principle of naturalness implies  that physics beyond the
standard model must be present at or below the ``'t Hooft
scale'' $4 \pi m_{\rm W}/g_w\sim 1 \TeV$ \cite{thooft}. In the next few
years several experiments will probe Flavor Changing
Neutral Currents (FCNC) and CP violation in the B system,
providing both new tests of the Standard Model (SM) and
potential clues to new physics  up to energies near $1000\TeV$.
These experiments may be the first
to provide evidence
for physics beyond the SM.  New physics
in rare decays of $B$ mesons and in
studies of CP violation in the $B_d$ and
$B_s$ systems can originate from: two non-SM phases
$\theta_{d,s}$ in the $\Delta B=2$ operators for $B_{d,s}$
mixing;
new phases in
the $\Delta B=1$ $b\to d$ and $b\to s$ hadronic transitions
(``penguins'');
disagreement between CP violation in the $B$ system and
$\epsilon$ in the kaon system;
or departure of  $\Delta m_{B_d}$ and/or $\Delta m_{B_s}$ from SM predictions.

In this Letter we show that all of the above effects
are likely to occur and may be
measurable in a class of theories recently proposed by three of us,
called \hbox{``Effective Supersymmetry'' \cite{CKN}.}  Effective
Supersymmetry is a new approach to the problem of naturalness in the
weak interactions, providing an experimentally acceptable suppression
of FCNC and electric dipole moments (EDMs) for the first two families
while avoiding fine tuning in the Higgs sector. In such a theory
nature is approximately supersymmetric above a scale $\tilde M$, with
$1\TeV\ll\tilde M\ltap 20\TeV$.  Unlike the Minimal Supersymmetric
Standard Model (MSSM) \cite{universality} however, most of the
superpartners have mass of order $\tilde M$ and only the Higgsinos,
gauginos, top squarks, and left handed bottom squarks need be lighter
than the 't Hooft scale. FCNC and EDMs for light quarks and leptons
are small even for large CP violating phases in supersymmetry breaking
parameters, due to approximate decoupling of the first two families of
squarks and sleptons.  Below $\tilde M$, the effective theory does not
appear supersymmetric, but is nevertheless natural, because of
substantial cancellations in quadratically divergent radiative
corrections.

The superpartner spectrum of Effective Supersymmetry can
result from new gauge interactions, which are
responsible for supersymmetry breaking and which couple more strongly to the
first two families than the top quark and up-type
Higgs. These new interactions could also explain the fermion
mass hierarchy and the absence of observed $B$ and $L$
violation.

We have computed the possible effects on B factory
physics from the light  gauginos, Higgsinos, and top and
bottom squarks.  We find different and larger effects are
possible than in the MSSM with squark universality
\cite{universality,mssmbphysics} or alignment
\cite{alignment}. Nonuniversal masses for the third generation of squarks
and sleptons have also been considered in \cite{DP,nonunithird}, and
the effects of nonuniversal masses and new phases 
for the third generation of squarks
on B physics has been considered previously in the context of grand
unified theories \cite{soten,desh}. 

B factory experiments will be able to
distinguish the effects of the standard model CKM
phases\cite{AKL}:
\begin{eqnarray}
\alpha\equiv\arg\left(-{V_{td}V^*_{tb}\over V_{ud}V^*_{ub}}\right)&&\quad
\beta\equiv\arg\left(-{V_{cd}V^*_{cb}\over V_{td}V^*_{tb}}\right) \nn\\
\gamma\equiv\arg\left(-{V_{ud}V^*_{ub}\over V_{cd}V^*_{cb}}\right)&&\quad
\gamma'\equiv\arg\left(-{V_{tb}V^*_{ub}\over V_{ts} V_{us}^*}\right) \nn\\
{\delta}\equiv\arg\left(-{V^*_{tb}V_{ts}\over  V^*_{cb}V_{cs}}\right)
&& \quad\omega\equiv\arg\left(-{V_{ud}V^*_{us}\over V_{cd}
V_{cs}^*}\right)\nn
\end{eqnarray}
from the effects of new physics (such as supersymmetric box and
penguin diagrams)
\cite{breviews}.
Note that with these definitions there are two identities,
\be
\alpha+\beta+\gamma=\pi; \quad \omega = \gamma-\gamma'-\delta \ .
\ee
{}From direct measurements of CKM parameters, and the assumption that
there are no new physics contributions to decay amplitudes which can
compete with SM tree level processes,  $|\omega|\ltap 0.2$.
Note however that $\omega > \CO(10^{-3})$ requires both CKM
non-unitarity {\em and} new physics in $K$--$\bar K$ mixing.
CKM unitarity also constrains $|{\delta}|< 0.03$.

We first consider  the effects of new
physics through $\Delta B=2$ operators.
Many of the  time dependent asymmetries resulting
from the interference between $B^0$--$\bar B^0$ mixing and
decay into CP eigenstates \cite{CSBS}
are cleanly predicted in the Standard Model
as a function of the  Cabibbo-Kobayashi-Maskawa (CKM) parameters
\cite{DR}. While  the direct decay amplitudes in table 1 will  be
dominated by
SM physics, the CP violating asymmetries
which result from interference between mixing and decay are
sensitive to  gauginos, Higgsinos, and squarks through box diagrams which
can produce nonstandard $\Delta B=2$ effects.
\begin{tabular}{|l|c|c|}
\hline
\hfil Decay \hfil &\hfil Quark Process \hfil &\hfil $A_{\rm
CP}$\hfil
\\ 
\hline
&&
\\ [-8pt]
$B_d^0 \rightarrow \pi^+ \pi^-$ &
$\bar{b} \rightarrow \bar{u}u\bar{d} $&
$ \sin{2 (\alpha - \theta_d)}$
\\
\hline
&&
\\ [-8pt]
$B_d^0 \rightarrow D^+ D^-$ &
$\bar{b} \rightarrow \bar{c}c\bar{d} $&
$- \sin{2(\beta + \theta_d)}$
\\ 
\hline
&&
\\ [-8pt]
$B_d^0 \rightarrow \psi K_s$ &
$\bar{b} \rightarrow \bar{c}c\bar{s} $&
$- \sin{2(\beta + \theta_d+\omega)}$
\\ 
\hline
&&
\\ [-8pt]
$B^\pm \rightarrow D_{\rm CP} K^\pm$ &
$\bar{b} \rightarrow \bar{c}u\bar{s},\bar{u}c\bar{s} $&
$\gamma - \omega\equiv $
\\
$B_d^0 \rightarrow D_{\rm CP}K^*$&
&
$\quad\gamma' +\delta$
\\
\hline
&&
\\ [-8pt]
$B_s^0 \rightarrow \psi \phi$ &
$\bar{b} \rightarrow \bar{c}c\bar{s} $&
$\sin{2(\delta - \theta_s)}$
\\ 
\hline
&&
\\ [-8pt]
$B_s^0 \rightarrow D_s^\pm K^*$ &
$\bar{b} \rightarrow \bar{c}u\bar{s},\bar{u}c\bar{s}$&
$\gamma' - \delta + 2 \theta_s$
\\ 
\hline
\end{tabular}
\vskip .2 in
Table 1. {\it CP asymmetries measured in B decays}
\vskip .2 in

\noindent
This new physics may be parameterized by two phases $\theta_d,\theta_s$:
\be \theta_{d,s}\equiv {1\over 2}\arg
\left({\langle  B_{d,s}|\CH_{\rm eff}^{\rm full}|\bar B_{d,s}\rangle\over
\langle  B_{d,s}|\CH_{\rm eff}^{\rm SM}|\bar B_{d,s}\rangle}\right)\ ,
 \ee
where $\CH_{\rm eff}^{\rm full}$ is the effective Hamiltonian
including both standard and SUSY contributions, and $\CH_{\rm
eff}^{\rm SM}$ only includes the effects of the standard model
box diagrams.

With these definitions, CP violating asymmetries in B processes
measure the angles as indicated in table 1. These processes
have been discussed in the SM  in \cite{decays}.
The measurements of $\alpha-\theta_d$ and $\beta+\theta_d$
are somewhat influenced by penguin contributions,
whose effects must be removed \cite{penguins}.
A subtle point is the presence of $\omega$ in $A_{\rm CP}$ for $B_d^0
\rightarrow \psi K_s$. This arises since we cannot assume the phase in
$K$--$\bar K$ mixing is given by the SM analysis \cite{desh}. 
However we do know,
since $\epsilon_K$ is small, that the phase is nearly the same as that
in $K$ {\it decay\/}, given by  $\arg{V_{ud}V^*_{us}}$.

Provided that penguin contributions to the decays of table 1 can be
removed, $\alpha, \beta, \theta_d, \omega$ and $\delta-\theta_s$
may be extracted from experiments \cite{desh} as indicated in figures 1 and 2.
With the additional assumption of CKM unitarity, $\delta$ is
quite small, and  $\theta_s$ may be extracted separately \cite{wolf}.

We can estimate the sizes of these effects by comparing the
superpartner contribution to
$\Delta B=2$
operators with the Standard Model.
Effective Supersymmetry
requires the squarks $\tilde Q_3$ and $\tilde{\bar T}$ to have masses
$\ltap 1\TeV$. These mass eigenstates are mixtures of flavor
eigenstates (where squark flavor, indicated by a lower case letter,
is defined by the gluino coupling to
the corresponding quark) \cite{CKN,prep}
\begin{eqnarray}\label{eq:zdefinitions}
\tilde Q_3\equiv\pmatrix{\tilde T\cr \tilde B}&\equiv& Z^{q}_{tT}
\pmatrix{ \tilde t\cr V_{tb}\tilde b+V_{ts}\tilde s+ V_{td} \tilde d}
\\&&+
 Z^{q}_{cT}
\pmatrix{ \tilde c\cr V_{cb}\tilde b+V_{cs}\tilde s+ V_{cd} \tilde d}
\nn\\&&+
 Z^{q}_{uT}
\pmatrix{ \tilde u\cr V_{ub}\tilde b+V_{us}\tilde s+ V_{ud} \tilde
d}\ ,\nn
\end{eqnarray} 
\be
{\tilde {\bar T}}\equiv Z^{\bar u}_{tT}\tilde{\bar t}
+Z^{\bar u}_{cT}\tilde{\bar c}
+Z^{\bar u}_{uT}\tilde{\bar u}\  .
\ee
Here $V$ is the CKM matrix, while the $Z$ factors arise 
\begin{figure}[t]
\centerline{\epsfxsize=3 in \epsfbox{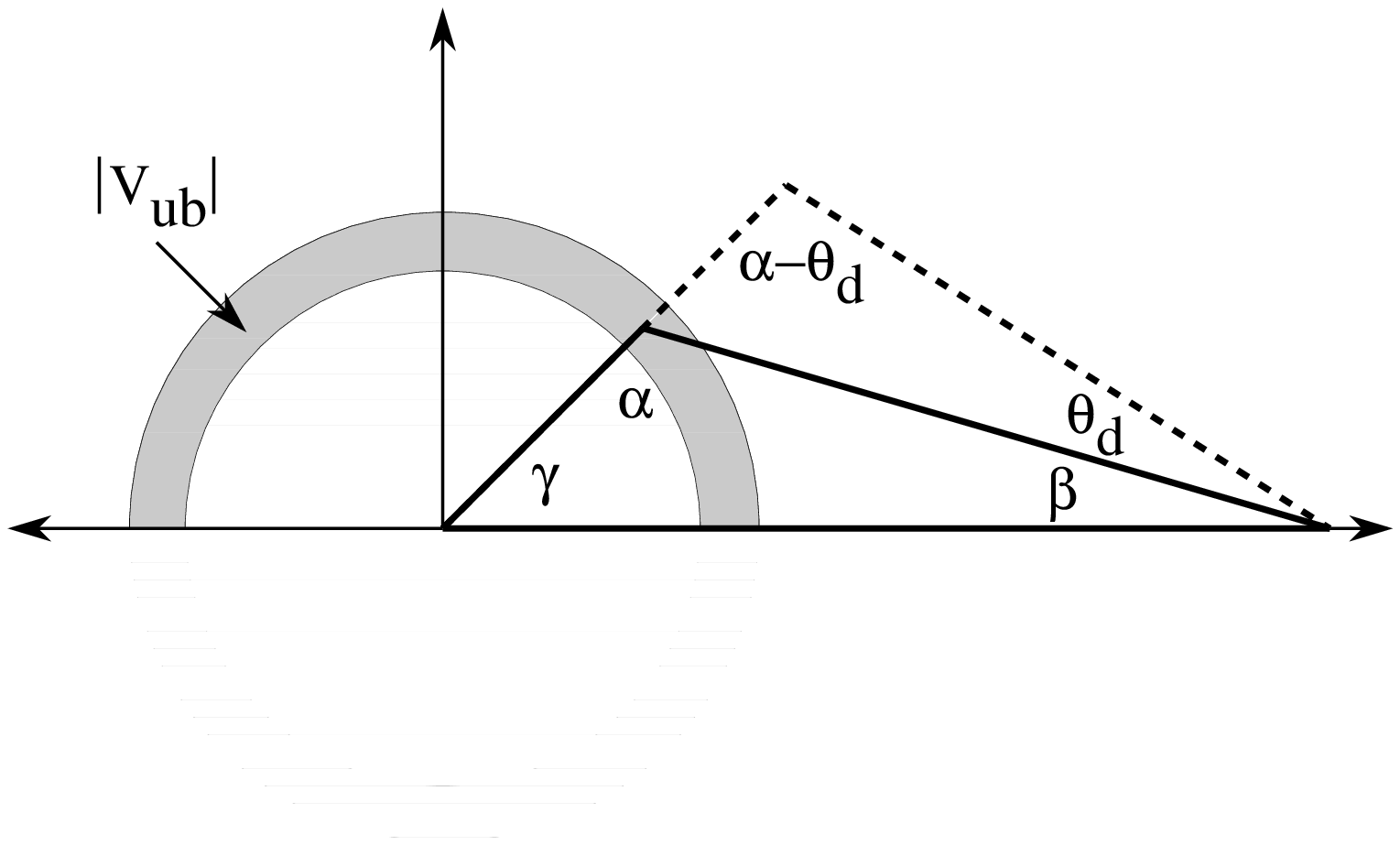}}
\vskip-.6in
\noindent
Fig 1. {\it Solid triangle corresponds to the CKM unitarity condition
$V_{ud} V_{ub}^* + V_{cd} V_{cb}^* + V_{td} V_{tb}^*=0$.  The angles
$(\alpha-\theta_d)$
and $(\beta+\theta_d)$ are measured; $\alpha, \beta$ and  $\theta_d$
may then be reconstructed from knowledge of $|V_{ub}|$.}
\end{figure}
\noindent from diagonalizing the squark mass matrix
in the quark mass eigenstate basis
(we neglect left-right squark mixing, which is small in realizations
of Effective Supersymmetry which have been studied to date \cite{CKN,DP}).
The $Z$ matrices satisfy $\sum_{i=u,c,t} |Z_{iT}^q|^2=1\ ,
\sum_{i=u,c,t} |Z_{iT}^{\bar u}|^2=1$.
Naturalness imposes order of magnitude constraints on the $Z$ factors:
to avoid fine tuning in the Higgs sector, we require
\be
|Z^{q}_{cT}|, |Z^{q}_{uT}|, |Z^{\bar u}_{cT}|,
|Z^{\bar u}_{uT}| \ltap {1\TeV\over\tilde M}\ .
\ee
while naturalness of the squark mass matrix requires
\cite{prep}:
\begin{eqnarray}\label{eq:natconstraint}
|Z^{q}_{uT}|& \ltap &\max\left({m_{\tilde Q_3}\over\tilde M},|V_{ub}|\right),\\
\ |Z^{\bar u}_{uT}|& \ltap&\max\left({m_{\tilde{\bar T}}\over\tilde M},
{m_{\tilde{Q}_3}\over\tilde M}\right),\nn
\end{eqnarray}
and similarly with $u$ replaced by $c$.

The box diagrams with left handed light squarks and gluinos give
\cite{HKT}\begin{eqnarray}\label{eq:llgluinobox}
\CH_{\rm eff}^{\tilde g}&=&
 {\alpha_s^2\over 36 m^2_{\tilde B }}
(Z_{dB}^{q}Z_{bB}^{q*})^2f_1(x_g)Q_1\\
&\approx&\left({6.4\cdot 10^{-12}\over \GeV^{2}}\right)
\left({1000 \GeV\over  m_{\tilde B }}\right)^2
\left({V_{td}+Z_{uT}^{q}\over 0.05}\right)^2Q_1,\nn
\end{eqnarray}
 where 
\begin{eqnarray}   
Q_1&=&\bar b^\alpha_L\gamma_\mu d_{\alpha L}
\bar b^\beta_L\gamma^\mu d_{\beta L}\nn \\
f_1(x)&=& {11 + 8\,x - 19\, x^2  + 26\,x\,\log (x) + 4\, x^2 \, 
\log(x)  \over  ( 1 - x )^3}\nn\\
\label{eq:fonedefinition}
Z^q_{q' B}&\equiv&\sum_{i=u,c,t}Z_{iT}V_{i q'},\qquad q'\equiv d,s,b\ .\nn
 \end{eqnarray}
and we have evaluated the function at $x_g\equiv m_{\tilde g}^2/ m_{\tilde B
}^2\simeq 0.1$.

Unless gluinos are significantly  heavier than squarks, charginos and
neutralinos (which does not occur in any realization of
effective supersymmetry  discussed in the literature  \cite{CKN,DP}),
box diagrams from chargino and neutralino exchange produce a
contribution suppressed by $\CO(\alpha_w/\alpha_s)^2\sim 0.1$ when compared
with the gluino boxes.
\begin{figure}[t]
\centerline{\epsfxsize=3 in\epsfbox{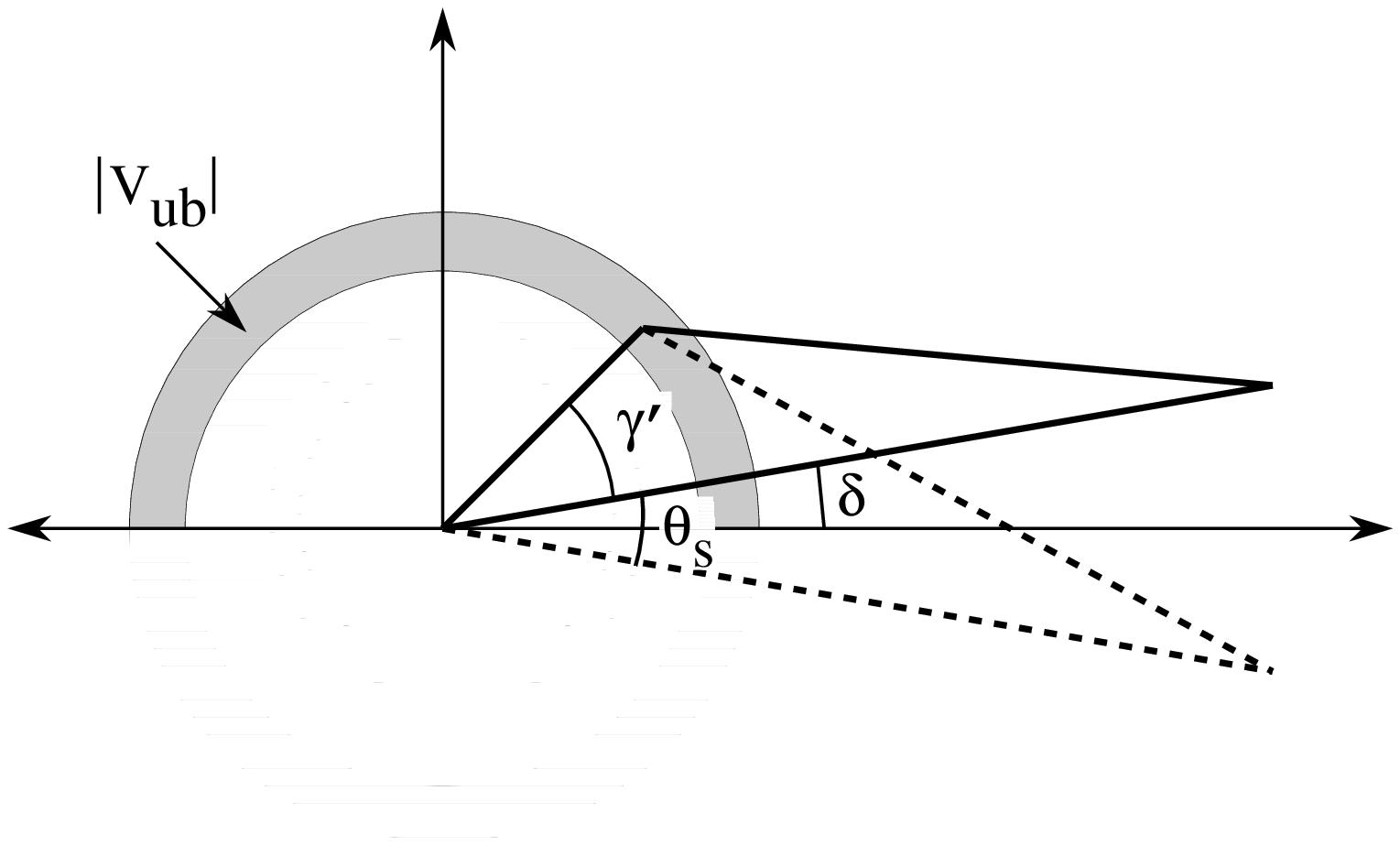}}
\vskip-.4in
\noindent
Fig 2. {\it Solid triangle corresponds to the CKM unitarity condition
$V_{tb} V_{ub}^* + V_{ts} V_{us}^* + V_{td} V_{ud}^* = 0$.  The angles
$(\gamma'+\delta)$ and
$(\delta - \theta_s)$ can be measured in $B_s$
decays while ${\delta}$ is constrained by CKM unitarity.
}
\end{figure}
\noindent Possible
exceptions are the charged Higgsino and charged
Higgs boxes which are proportional to $\lambda_t^4$.
However these have the same phase as the standard model
contribution.

{}From eq.~\ref{eq:llgluinobox} we see that even TeV mass squarks can
produce an order one effect on $B_d$--$\bar B_d$ mixing, detectable
via a $\theta_d$ as large as $\pm \pi/2$, or via a ratio for $x_s/x_d$
(where $x_{s,d}\equiv \Delta m_{B_{s,d}}/\Gamma_{B_{s,d}}$) which is
well outside the SM range.  For $B_s$--$\bar B_s$ mixing the effects
of the superpartner box diagrams can only be comparable to the SM
contribution for rather light ($\sim 200 \GeV$) $b$ squarks and
gluinos. A measurement of $\theta_s$ larger than 0.2 would suggest
that gluinos and a squark are lighter than $\sim 400$ GeV.

In the SM $\epsilon_K$ significantly constrains the CKM matrix.
However $\epsilon_K$ could be dominated by the contribution from
supersymmetric particles, even if all superpartners are as heavy as
$500 \TeV$. With $\sim 20$ TeV masses for the first two families of
squarks and with  susy mixing angles for the first two generations
squarks of order the Cabbibbo angle, the CP violating susy phases in
the down and strange squark couplings must be less than $\CO(1/30)$ or
the kaon CP violating parameter $\epsilon_K$ would be too large
\cite{CKN}. Note that suppressing this susy contribution to $\epsilon_K$ does
not preclude observing new CP
violating phases in B physics. However an interesting possibility 
is  that an approximate
CP symmetry renders all phases (including CKM phases) small. In this
case the CP violating asymmetries in $B$ decays would all be too small
to be easily measured.

In either the MSSM or in effective supersymmetry 
it is possible that $\Delta m_{B_d}$ could receive a
significant supersymmetric contribution which has the same phase
as the SM contribution.  Thus the values of
$\alpha,\beta$ determined by B physics could disagree with the values in
the SM given by $V_{ub},\Delta m_{B_d}$ and $\epsilon_K$, even if
$\theta_{d,s}$ are too small to measure.

Supersymmetry may also have significant effects through $\Delta B = 1$
operators.
Contributions to both
the $b\rightarrow d$ and $b\rightarrow s$ penguins can be
comparable to that of the SM but with different phases, 
provided gluino and
third family squark masses are lighter than $200 \GeV$. The SM
predictions for penguin operators, and methods for extracting
their effects from CP asymmetries has been extensively
discussed \cite{penguins,desh,wolf,GW}. In the  standard
model there is a large uncertainty in the phase of the $b\rightarrow
d$ penguin, however  the uncertainty in the phase of
the $b\rightarrow s$ penguins is of order $\delta$ if the three by
three CKM matrix is unitary.  Thus one can  search for  new CP
violating phases in penguin
contributions via, {\it e.g.}, the CP asymmetry in $B_d(\bar
B_d)\rightarrow\phi K_S$. 

Box and electroweak penguin diagrams involving superpartners can
affect the rates, 
polarizations, and lepton momentum distributions in $b\rightarrow (s,d)
\ell^+ \ell^-$, which can also be tested in B factories.  In the MSSM with
universality, the only  potential discrepancies larger than 5\% arise through
changes in the
coefficient $\CC_7$ \cite{bsell} in the effective Lagrangian (we follow the
notation of \cite{buras}). In Effective Supersymmetry with small left-right
squark mixing and heavy charged Higgs  the corrections to $\CC_7$
are small.
With a bottom squark lighter than $\sim 100\GeV$ and gluino lighter than $\sim
200\GeV$ it is possible to change the size and/or phase of the coefficient
$\CC_9$ by as much as 30\%. If the bottom and/or top squarks, the weak gauginos
and the $\tau$ charged slepton and/or $\tau$ sneutrino  have masses $\sim 100$
GeV, it is possible for box diagrams to change the size and phase of
$\CC_{9,10}$ (for the $\tau$ lepton only) by a maximum of $\CO$(10\%).

The B factories will also search for
mixing and CP violation in the $D^0$ system, which are both predicted to be
very small in the SM ($x_D\equiv \Delta
m_{D^0}/\Gamma_{D^0}\sim 10^{-4}$--$10^{-5}$, $y_D\equiv \Delta
\Gamma_{D^0}/(2 \Gamma_{D^0})\sim 10^{-2}$--$10^{-4}$,
$\epsilon_D\sim 10^{-4}$--$10^{-6}$)
\cite{ddbar}.
In Effective Supersymmetry there  can be
significant contributions to $x_D$ from both heavy squarks with masses
$\sim\tilde M$ and from the lighter third family squarks,  with comparable
maximum possible size.
For example the box diagrams  with a right handed top squark and
gluinos give a contribution
\begin{eqnarray}\label{eq:xd}
x_D&=& {\alpha_s^2M_DB_D f^2_D\over 54
m^2_{\tilde{\bar T}}\Gamma_D}
|(Z_{uT}^{\bar u}Z_{cT}^{\bar u})|^2f_1(x_g)\\
& \approx &  5\cdot10^{-4}
\left({1000\GeV\over m^2_{\tilde{\bar T}}}\right)^2
\left({f_D\sqrt{B_D}\over 200\, {\rm MeV}}\right)^2
\left({Z_{uT}^{\bar u}Z_{cT}^{\bar u}\over 0.0025}\right)^2\,\nn
\end{eqnarray}
where again we have taken $x_g \simeq 0.1$.
The current experimental bound is ($x_D<0.09$) \cite{PDG}.
Charm decays will be dominated by the SM contribution and so there are
no significant new contributions to $y_D$. We conclude
that unless suppressed by flavor symmetries, $D^0$--$\bar D^0$ mixing
could be much larger than in the SM, although substantially
smaller than the current experimental bounds.
The superpartner contribution may also  have a different phase
than the SM contribution. If
$\Delta m_{D^0}$ and ${\Delta\Gamma}/2$
turn out to be comparable,  $\epsilon_D$ could  be $\CO(1)$,
although $\epsilon_D$ is
difficult to measure if
$D^0$--$\bar D^0$ mixing is very slow.
In principle $D^0$--$\bar D^0$ mixing affects the
extraction of the CKM parameter $\gamma-\omega$ from $B\rightarrow
D_{\rm CP} K$ 
decays; however such effects are suppressed by $x_D,y_D$, and are
negligible. 
However even if $\epsilon_D$ is small,
$x_D$ may be as large as  $\CO(10^{-2})$, and then
CP violation in interference between $D^0$ mixing and decays
might be detectable \cite{BSN}.

In summary, Effective Supersymmetry, with naturalness and with $\tilde M\sim 20
\TeV$, allows for interesting new physics for B factories. Observable
possibilities which are precluded in other supersymmetric models
(assuming R-parity conservation)
include
 large values for the new physics parameters $\theta_d$ and
 $\theta_s$, and  large new phases in $b\rightarrow s$ penguins. 
$D^0$--$\bar D^0$ mixing is likely to be much larger than in the
standard model but very difficult to observe. 
Note that observation of large $\theta_s$, non-standard phases in
$b\rightarrow s$ penguins, or
measurable deviation from the SM
in $b\rightarrow (d,s) \ell^+\ell^-$, would  imply that gluinos
and third family squarks are lighter than $\sim 200$ GeV, {\it i.e.} 
within near term experimental reach.
Effective supersymmetry shares with other supersymmetric models the
possibility of nonstandard contributions to $\epsilon_K$ and
$B_d$--$\bar B_d$ mixing. 
 
\acknowledgements

A.C. was supported in part by the DOE under grant
\#DE-FG02-91ER40676. D.K. and F.L. were supported in part by DOE grant
DOE-ER-40561, and NSF Presidential Young Investigator award
\pyidk. A.N. was supported in part by the DOE under grant
\#DE-FG03-96ER40956. We thank Y.Nir for   useful correspondance.

\end{document}